\documentstyle[aps]{revtex}
\begin{document}

\title{Geometric statistical inference}
\author{Vipul Periwal}
\address{Department of Physics,
Princeton University,
Princeton, New Jersey 08544}

\def\dd{\hbox{d}}
\def\tr{\hbox{tr}}\def\Tr{\hbox{Tr}}
\def\ee#1{{\rm e}^{{#1}}}
%\baselineskip=14truept
\maketitle
\begin{abstract} 
Finite sample size corrections to 
the reparametrization-invariant  solution of the 
inverse problem of probability are computed, and shown to converge 
uniformly to the correct distribution. %A detailed geometric solution of the
%statistical inference problem in higher dimensions is given.

\end{abstract}

\section{Introduction}
A basic problem in statistics, with applications in divers 
fields, is the 
determination of  the probability distribution that underlies a finite set
of experimental results involving continuous variables.  From the
practical point of view, one usually has a definite finite-parameter
model for the 
probability distribution on theoretical grounds, and the parameters 
are fixed by fitting to the data set.  As is obvious, this 
parametric inference 
introduces a theoretical bias, since the true distribution may not 
even lie in the parameter set chosen.  
It is therefore of interest to 
attempt a direct determination of the probability 
distribution without resorting to finite-dimensional approximations.  
Of course, for any finite data set, one obtains only a probabilistic 
description of the probability distribution, but the spread of 
possible probability distributions decreases as the size of the data 
set is increased\cite{history}.  

This set-up is fundamental in pattern theory, a context in which 
reparametrization invariance is 
of great importance. In visual information processing algorithms, for 
example, if the 
inferred pattern depends on, say, the orientation of the object, then 
the object recognition capabilities of the algorithm are not likely 
to be efficient.  In speech recognition, variations in the tempo of
speech should, again, be treated as reparametrizations of 
the same underlying pattern, as should frequency variations. 

\def\prob{{\rm prob}}
In the Bayesian approach to the problem,  one starts by assuming a 
space of probability distributions, $\{w\},$ within which the true distribution 
lies, and given a set of observations, $f$, one obtains a probabilistic 
description of the true distribution by using 
\begin{equation}
\prob(w|f) = {{\prob(f|w)\cdot \prob(w)}\over \prob(f)};
\end{equation}
in words, the probability of the distribution $w$ being the true 
distribution, given $f,$ is the probability of $f$ given the 
distribution $w,$ multiplied by the probability of $w$ occuring in the 
space of all distributions, normalized by the probability of $f$ 
occuring in {\it any} of the distributions in the set $\{w\}.$
The {\it maximum likelihood} (ML) estimate of $w$ is then that distribution
$w$ that maximizes $\prob(f|w)\cdot \prob(w).$  To go further in this 
direction, one needs to figure out what the {\it a priori} 
probability $\prob(w)$ should be, and how to compute the ML estimate.
Clearly, to avoid bias, one wants the space of distributions to be as 
general as possible, consistent with computability.  
  
In one dimension, Bialek, Callan and Strong\cite{bialek}
have recently given an elegant formulation of
this problem for continuous distributions in one dimension. 
They used Bayes' rule to write the 
probability of the probability distribution $Q,$ given the data 
$\{x_{i}\},$ as
\begin{eqnarray}
P[Q(x)| x_1 , x_2 , ... , x_N]
\,
 = 
{{P[x_1 , x_2 , ... , x_N | Q(x)] P[Q(x) ]}
\over {P(x_1 , x_2 , ... , x_N )}} \, =
{{Q(x_1 ) Q(x_2 ) \cdots Q(x_N) P[Q(x)]}
\over{\int [dQ(x)] Q(x_1 ) Q(x_2 ) \cdots Q(x_N) P[Q( x)]}} ,
\label{conditional}
\end{eqnarray}
where the factors $Q(x_{i})$ arise because each $x_{i}$ is chosen 
independently from the distribution $Q(x),$ and $P[Q]$ encodes the 
{\it a priori} hypotheses about the form of $Q.$     The optimal 
least-square estimate of $Q,$ $Q_{{\rm est}}(x,\{x_{i}\}),$ is then
\begin{equation}
Q_{\rm est} (x; \{x_i\} )
=
{{\langle Q(x) Q(x_1 ) Q(x_2 ) \cdots Q(x_N) \rangle ^{(0)}}\over
{\langle Q(x_1 ) Q(x_2 ) \cdots Q(x_N) \rangle^{(0)}}} ,
\label{est}
\end{equation}
where $\langle \cdots \rangle^{(0)}$ denotes an  expectation value with
respect to the {\it a priori} distribution $P[Q(x)]$.  

In this field-theoretic setting, Ref.~\cite{bialek} assumed that 
the prior distribution $P[Q]$ should penalize large gradients, so
written in terms of an unconstrained variable $\phi \equiv  \ln (\ell 
Q) \in (-\infty,+\infty),$ they assumed 
\begin{equation}
P_\ell[\phi(x)]
 =  {1\over Z} \exp\left[ -{\ell\over2} \int dx (\partial_x \phi)^2
\right]
\, \times
\delta \left[ 1 - {1\over \ell} \int dx \ee{\phi(x)}\right] ,
\label{bcs}
\end{equation}
with $\ell$ a {\it global} length 
parameter that they later fixed by means of other 
considerations.  
This form for the prior distribution is very simple,  and quite 
minimal in terms of underlying assumptions, so 
is almost an ideal example of the Bayesian set-up.

There is, however, 
an important aspect in which this prior distribution does not measure 
up---a probability distribution is a density, and hence transforms in
a specific manner under reparametrizations of the data,
{\it e.g.} given a set of $N$ observations $x_{i},$ we
want to infer a distribution $w_{x}$ that is related in a very simple 
way to the distribution $w_{{\exp}}$ inferred from $\exp(x_{i}):$
\begin{equation}
w_{x}(z) = \exp(z)\cdot w_{\exp}(\exp(z)).
\end{equation}
This reparametrization covariance of the estimated distribution turns 
out to be a strong constraint on the distribution $\prob(w),$ and 
on the actual computation of the   estimated distribution.  In recent work, I 
found a geometric modification of the approach of Ref.~\cite{bialek},
which satisfies this desired reparametrization covariance\cite{me}.  
It is the details of this solution,  especially the corrections for
finite sample size (which are obviously of paramount importance for
practical applications), % and its analogue in higher dimensions,
that are explained  in the present paper.
 
This paper is organized as follows: Section 2 contains a short review 
of Ref.~\cite{me}, to keep the presentation here relatively 
self-contained.  In Section 3, the one-dimensional solution  for finite
sample sizes is given, along with a discussion of corrections arising 
from fluctuations about the saddlepoint distribution.  %In Section 4, 
%after summarizing some facts about conformally Euclidean geometry, I 
%give the detailed solution of the inference problem in dimensions 
%larger than 3. 
In section 4, I present some conclusions, with comments on the 
solution of the inference problem in dimensions larger than 3.
 
\section{Review}

Instead of the form (eq.~\ref{bcs}) used in Ref.~\cite{bialek}, 
I write $Q(x) \equiv \sqrt{h(x)} \exp(\phi(x)) ,$ 
with $h(x)$ a metric in one dimension.  Hence $\sqrt{h(x)}$ is a 
scalar density, and so is $Q,$ which is crucial for a
reparametrization covariant solution.  An intuition for the r\^ole that
$h$ plays is that it is a local analogue of $\ell,$ determining binning
intervals locally.  Set 
\begin{eqnarray}
P[\phi,h] \equiv {1\over Z} \exp\left[ -{\ell\over2} \int dx 
\sqrt{h(x)}^{{-1}}(\partial_x \phi)^2
\right] \delta \left[ 1 - {1\over \ell} \int dx \sqrt{h(x)} \ee{\phi(x)}\right] ,
\label{PP}
\end{eqnarray}
The inverse of the metric is just $1/h(x)$ and the reparametrization  
invariant volume element is $\sqrt{h(x)} dx,$ so $P[\phi,h]$ is 
reparametrization invariant.  Now, we want to evaluate
\begin{eqnarray}
&&\langle Q(x_1 ) Q(x_2 ) \cdots Q(x_N) \rangle^{(0)}
\nonumber\\
&&\,\,\,\,\,
=
\int D\phi {{ Dh}\over{Diff_{+}}}
P[\phi,h] \prod_{i=1}^N \sqrt{h} \exp[\phi(x_i)] \\
&&\,\,\,\,\, =
 {1\over Z} \int {{d\lambda }\over{2\pi}}
\int D\phi {{ Dh}\over{Diff_{+}}}
\exp\left[ - S(\phi,h; \lambda ) \right] ,
\label{problem}
\end{eqnarray}
where 
\begin{eqnarray}
S(\phi,h; \lambda ) =
{1\over2} \int dx \sqrt{h(x)}^{{-1}}(\partial_x \phi)^2 
\,
+i{\lambda } \int dx \sqrt{h(x)} \ee{\phi(x)}
-\sum_{i=1}^N [\phi (x_i )+ {1\over 2} \ln h(x_{i})] - i\lambda.
\label{accion}
\end{eqnarray}
Notice that the integral over all metrics has been divided by 
the volume of the group of orientation preserving diffeomorphisms.  In one 
dimension, this division eliminates all but one global degree of freedom from
the metric.  There 
is no operational way to distinguish between the factor 
$\sqrt{h(x_{i})}$ and $\exp(\phi(x_{i}))$---these 
must occur together in $Q(x_{i}).$  Taking the local symmetry into account, 
the number of local degrees of freedom is the same as in the approach 
of Bialek et al.\cite{bialek}.

The equations of motion that follow from varying $S$ are
\begin{eqnarray} 
&&-{1\over 2} (\phi')^{2} {1\over h} + i\lambda \exp(\phi) -
\sum_{i} {1\over \sqrt{h(x_{i})}} \delta(x-x_{i}) = 0 \ ,
\label{hvar}
\\
&& - \left({1\over \sqrt{h}}\phi'\right)' 
+i\lambda\sqrt{h}\exp(\phi) -\sum_{i}\delta(x-x_{i}) = 0 \ ,
\label{pvar}
\\
&& \int dx \sqrt{h} \exp(\phi) =1\ ,
\label{lvar}
\end{eqnarray}
where I have used primes to denote $d/dx.$    $\delta(x)$ denotes the
scalar density such that $\int dx \delta(x) =1.$
Introduce a variable 
\begin{equation}
y(x) = \int^{x} \sqrt{h(s)} ds,
\label{careful}
\end{equation}
then 
it follows from eq.~\ref{hvar} and eq.~\ref{pvar} that 
\begin{equation}
{d^{2}\phi\over{dy^{2}}} =   {1\over 2} \left({d\phi\over 
dy}\right)^{2}.
\end{equation}
Note that the use of eq.~\ref{careful} is limited by the 
non-singularity of $\sqrt{h}.$

It is now necessary to be careful about the limits of integration.  
This care in the boundary terms is 
unnecessary at $N=\infty,$ in accord with the fact that any finite 
data set will   not indicate the true limits of the probability 
distribution.
Suppose that $x$ ranges from $x_{-}$ to $x_{+},$ then 
integrating  eq.~\ref{pvar} I find 
\begin{equation}
N - i\lambda + \int_{x_{-}}^{x_{+}}
dx ({1\over \sqrt{h}}\phi')' =0 \ ,
\end{equation} 
so, using eq.~\ref{lvar},
%\begin{equation}
%\int_{y_{-}\equiv y(x_{-})}^{y_{+}\equiv y(x_{+})} dz \exp(\phi(z)) = 1\ ,
%\end{equation}
it follows that 
\begin{equation}
\exp(-\phi) = (y_{+}-y_{-}) {(y-c)^{2}\over(y_{-}-c)(y_{+}-c)}
\label{py}
\end{equation}
where $c$ is an arbitrary constant of integration.  $c$ is restricted 
only by $c>y_{+}>y_{-},$ or $y_{+}>y_{-}>c,$  since $\exp(\phi)$ must
be positive.  This `cyclic' constraint on $y_{+},y_{-},c$ arises 
because the algebra of
infinitesimal reparametrizations of the line has a subalgebra 
isomorphic to $sl(2,{\bf R}).$  

In order to determine $y(x),$ which satisfies
\begin{equation}
{(y_{+}-c)(y-y_{-})\over(y_{+}-y_{-})(y-c)} = \int_{x_{-}}^{x} 
\left[{1\over N} \sum_{i}\delta(x'-x_{i})\right] dx'\ ,
\label{singular}
\end{equation}
observe that the cross ratio on the left is projectively invariant, 
{\it i.e.}, invariant under transformations of the form 
\begin{equation}
z \mapsto {{\alpha z+\beta}\over {\gamma z+\delta}},
\end{equation}
with $\alpha,\beta,\gamma,\delta $ real, $\alpha\delta-\beta\gamma=1.$
This amounts to a three-parameter family of
equivalent solutions $y(x)$ determined by the data.  
This projective invariance can be fixed by setting $c=\infty,y_{+}=1$
and $y_{-}=0,$ which implies  $\phi=0.$  

Operationally, this means 
that the next measured data point will be observed with equal 
probability at any value of $y$ in the interval $[0,1].$  An important 
point to notice about this solution is that it is only valid at 
$N=\infty.$  At finite $N,$ the metric determined by this solution is 
not smooth, being a sum over $\delta$ functions, hence we must take 
more care in solving eq.'s~\ref{hvar},\ref{pvar},\ref{lvar} for finite $N.$ 
Indeed, the action evaluated at
this  solution is infinite for finite $N.$  
\section{Finite $N$ corrections}

\def\part{\partial}
\def\al{\alpha}
\def\be{\beta}
\def\eps{\epsilon}
\def\lam{\lambda}
\def\ga{\gamma}
\def\del{\delta}
Obviously, for any practical application of this theory, one needs a 
systematic scheme for computing finite sample size corrections.  
To understand the theory at finite $N,$ and to make contact with the 
higher-dimensional
theory commented on in Section 4,  we rewrite eq.~\ref{accion} as 
\begin{equation}
S = {1\over 2} \int dy (\part_{y}\phi)^{2} +i\lam \left(\int dy  
\ee{\phi} - 
1\right)
+N\left[-\int dy \hat P_{y}(y) \phi(y) - \int dx \hat P_{x} \ln 
\sqrt{h(x)}\right],
\end{equation}
where $y(x)$ is as in  eq.~\ref{careful}, $\hat P_{x}(x) \equiv {1\over N} \sum 
\delta(x-x_{i}),$ and $\hat P_{y}(y) y' \equiv 
\hat P_{x}(x).$  It is clear in this form  that there are two 
independent functional variations of $S,$ one with respect to 
$\phi(y),$ and the other with respect to $y(x):$
\begin{eqnarray}
&& - \part_{y}^{2}\phi +i\lam \exp(\phi) -N\hat P_{y}(y) = 0 \ ,
\label{acc1}
\\
&& \part_{x} \left[ {{\hat P_{x}(x)} \over y'(x)}\right] = 0 \ .
\label{acc2}
\end{eqnarray}

For $N$ large, suppose $\hat P_{x}(x) = P(x) + {1\over {\sqrt N}} \rho(x),$
with $P(x)$ a continuous density and $\rho(x): \langle 
\rho(x)\rho(x')\rangle = P(x)\delta(x-x'),$ following 
Ref.~\cite{bialek}.  Now, a solution to eq.'s~\ref{acc1} and 
\ref{acc2} at leading order in $N$ is $\phi_{0}=0,$ and $\sqrt{h}(x) 
\equiv \part_{x}y_{0}(x) = 
P(x),$  as reviewed in Sect.~2.  Indeed, it would seem from 
eq.~\ref{singular} that this solution is exact, and no modification 
for
finite $N$ is necessary.  This is {\it not} true, since for finite sample 
sizes eq.~\ref{singular} defines a singular metric, and hence is  
inconsistent with our calculations which have assumed that the metric 
is non-singular implicitly. 

We can, however, solve eq.~\ref{pvar} 
to
get $\phi = \phi\left[\sqrt{h}, \hat P_{x}\right],$ and then use this 
solution in eq.~\ref{hvar} to determine $h.$  We shall find that the 
leading correction at finite $N$ is a correction to $\phi$ at 
$O(N^{{-1/2}}),$ with no correction to $\sqrt{h} = P(x).$
This ordering of the computational steps parallels the
calculation in \cite{bialek}, with the determination of $h$ here the 
analogue of the determination of the length scale in \cite{bialek}. 
What is different in the present approach, is that  the formul\ae\ 
derived here are reparametrization-covariant, {\it e.g.}, square roots of
densities do not appear. 
To leading order, we find $i\lam=N,$ and 
\begin{equation}
\exp(\phi_{0}) = {P(x)\over\sqrt{h}}.
\label{absolute}
\end{equation}
Notice that eq.~\ref{absolute} has the correct 
reparametrization covariance by construction, with a function equated 
to a ratio of densities.

At the next order, expanding eq.~\ref{pvar} about $\phi_{0}, 
 i\lam=N,$ we find  
\begin{equation}
\left[- {1\over \sqrt{h}(x)}\part_{x}{1\over\sqrt{h}(x)}\part_{x} 
+N\ee{\phi_{0}}\right]\phi_{1} \equiv \left(-\Delta_{h} + N 
\ee{\phi_{0}}\right)\phi_{1} = 
 \left[\sqrt N {\rho(x)\over 
\sqrt{h}(x)} + \Delta_{h}\phi_{0}\right] \ .
\label{lapdog}
\end{equation}
Eq.~\ref{lapdog}  is easily solved to obtain
\begin{equation}
\phi_{1}(x) = \int dx'\sqrt{h}(x') \left[\sqrt{N}{\rho(x')\over \sqrt{h}(x')} + 
\Delta_{h}\phi_{0}(x')\right] K(x,x'), 
\label{solute} 
\end{equation} 
with the leading term in $K$ given by 
\begin{equation}
 K(x,x') \approx 
{1\over 2\sqrt{N}} 
\left[\ee{\phi_{0}}(x)\ee{\phi_{0}}(x')\right]^{{-1/4}} 
\exp\left(-\sqrt{N}\int^{\max(x,x')}_{\min(x,x')}\sqrt{h}
\ee{\phi_{0}/2} dx''\right)\ .
\end{equation}
Note the reparametrization invariance 
of eq.~\ref{solute}.

We can now insert eq.'s~\ref{absolute},\ref{solute} in 
eq.~\ref{accion}, 
and vary with respect to $h$ to obtain $h=h(\hat P_{x}).$  This order 
of solving the saddlepoint equations is somewhat simpler than solving
the equations obtained by independent variations---the results are,
obviously, independent of this order  of solution.
We find
\begin{equation}
S_{{red}}[h] = -N \int dx \left[P + {1\over \sqrt{N}}\rho\right] \ln P +
{1\over 2} \int \left(NP\phi_{1}\left[\ln\sqrt{h}-\ln P 
-\phi_{1}\right]
-\sqrt{N} \rho \left[\ln \sqrt{h} +\phi_{1}- \ln P\right]\right) \ ,
\label{reduced}
\end{equation}
with $\phi_{1}=\phi_{1}(h).$
Now, it is important to note that while $S_{red}$ appears to include  
terms with   powers of $N $ that should be ignored at first glance,
the variation of $\phi_{1}$ with 
respect to $h$ is $O(1)$ even though $\phi_{1}=O(N^{{-1/2}}).$  Thus, 
varying eq.~\ref{reduced} with respect to $h,$ we find
\[ {1\over 2} NP {{\delta\phi_{1}}\over{\delta\sqrt h}}\left(\ln\sqrt 
h - \ln P\right) +  O(\sqrt N) = 0 \]
so to leading order in $N,$ 
 $\sqrt{h_{0}}=P, \phi_{0}=0$  so our old result
is recovered.  
However, at the next order, we find {\it no} correction to
$h$ of $O(N^{{-1/2}}).$  Thus the leading finite size correction is
entirely encoded in eq.~\ref{solute}, with $\sqrt h = P.$  At higher 
orders in $N^{{-1/2}},$ there are, of course, further corrections to 
$\phi,$ and to $h.$

Consider now the computation of the functional integral expanded 
about the saddlepoint solution found above.  The division by the 
volume of the group of diffeomorphisms eliminates the integration over
fluctuations in the metric, so we are left with computing the 
integral over the fluctuations in $\phi.$ This integration gives the 
operator determinant  
$\det^{-1/2}(-\Delta_{h} + N\ee{\phi_{1}}).$   This determinant can be 
computed in the large $N$ limit by the standard van Vleck technique,  
explained in Ref.~\cite{coleman}, for example.
Thus
\begin{equation}
\det\left(-\Delta_{P^{2}} + N\ee{\phi_{1}} \right)^{{-1/2}} \propto
\exp\left(-{\sqrt{N}\over 2}\int dx 
P(x) \ee{\phi_{1}/2} \right) \ ,
\label{det}
\end{equation}
which differs  from the corresponding expression in 
Ref.~\cite{bialek}. 
Eq.~\ref{det} is reparametrization invariant.
Putting everything together, eq.~\ref{problem} is
given by 
\begin{equation}
\prod_{i=1}^{N} P(x_{i}) \exp\left(-{\sqrt{N}\over 2}\int dx 
P(x) \ee{\phi_{1}/2} -{1\over 2}\int dx  
P(x)^{-1}(\part_{x}\phi_{1})^{2}  \right) \ .
\label{finale}
\end{equation}
Eq.~\ref{finale} can be used to evaluate the least-square estimate of 
the inferred distribution.

Finally, we compute the magnitude of the correction $\phi_{1}:$
%\begin{equation}
%\langle \phi_{1}(x_{1})\phi_{1}(x_{2})\rangle_{\rm connected} 
%= {1\over 4} \ee{-\sqrt 
%N \int^{\max(x_{1},x_{2})}_{\min(x_{1},x_{2})} dx' P(x')}
%\left[
%\int_{{x_{-}}}^{\min(x_{1},x_{2})} dx_{3}\ee{-2\sqrt 
%N \int_{x_{3}}^{\min(x_{1},x_{2})} dx' P(x')} +
%\int^{\max(x_{1},x_{2})}_{\min(x_{1},x_{2})}P(x_{3}) dx_{3}
%+ \int^{{x_{+}}}_{\max(x_{1},x_{2})} dx_{3}\ee{-2\sqrt 
%N \int^{x_{3}}_{\max(x_{1},x_{2})} dx' P(x')} 
%\right]\ ,
%\end{equation}
%which implies, in particular,
\begin{equation}
\langle\phi_{1}(x)^{2}\rangle_{\rm connected} \sim {1\over 4\sqrt{N}}\ ,
\label{slick}
\end{equation}
independent of $x,$ implying {\it uniform} convergence  
to the solution reviewed in Sect.~II.  This uniform convergence is of
great practical importance, since it implies that our estimated 
distribution will make sense even in regions where $P$ is small.  
The uniformity can be directly traced to the reparametrization 
invariance of our computations, which enables the binning size to be
adjusted locally enabling a  more accurate determination of the distribution.
As this uniformity is of great importance for practical applications, 
it is all the more gratifying that a re-analysis motivated by 
conceptual concerns regarding reparametrization covariance 
leads to a computational improvement.

\section{Conclusions}

The statistical inference problem is of great interest in biophysics, 
for example, when $x$ is a vector variable.  Bialek, Callan and 
Strong\cite{bialek} gave an account of a higher dimensional variant of
their theory, but I want to add a few remarks  here regarding 
 a geometric version
sketched in \cite{me}, taking 
into account reparametrization invariance.   In Sect.'s II and III, we 
saw that a reparametrization-covariant Bayesian set-up could be 
obtained by 
introducing a metric, {\it and} a scalar field, dividing out by
the group of diffeomorphisms.  Since a metric in one dimension has 
only one global degree of freedom, this left only one local scalar degree 
of freedom, which is what we expect to need  to fit the data, since the 
data determines a scalar density on the interval $[x_{-},x_{+}].$ 

In higher dimensions, the aim is still to determine a scalar 
density, but the number of local degrees of freedom in the metric in $d$
dimensions is $d(d+1)/2$ and reparametrizations form a $d$ parameter
local symmetry group.  It is clear then that a geometric formulation 
will require a reparametrization invariant constraint to reduce the 
number of degrees of freedom in the metric, to the required $d+1$
local degrees of freedom.

Reparametrization invariant constraints must be given by 
the vanishing of tensor quantities, so in this case we have to 
construct appropriate tensors from the metric.  Investigating the 
constraints provided by the vanishing of various curvature tensors, 
since we expect only a scalar local degree of freedom to 
survive after imposing the constraint  and dividing out by 
reparametrizations, we are naturally led to consider conformally 
Euclidean metrics, which can indeed be characterized by the vanishing 
of a tensor constructed from the curvature, the Weyl curvature, 
$W_{{\alpha\beta\gamma}}^{\delta}.$
$W$ vanishes identically in $d\le 3,$ so the geometric theory  
will only be valid for $d>3.$  

The proposed  generalization of eq.~\ref{PP} to higher dimensions is 
now given by 
\begin{equation}
P[h_{\alpha\beta}] \propto  \exp\left(-\int d^{d}x \sqrt{h}
{\cal L}\right)  \delta\left(1-\int d^{d}x 
\sqrt{h(x)}\right) 
\prod_{x}\delta\left(W_{\alpha\beta\gamma}{}^{\delta}(x)\right)
\label{d3first}
\end{equation}
where ${\cal L}$ is a scalar constructed out of the
metric $h_{\alpha\beta},$ constrained to be a metric of vanishing 
Weyl curvature, $W_{\alpha\beta\gamma}^{\delta},$ 
and $\sqrt{h}\equiv \det^{1/2}(h_{\alpha\beta})$ as 
usual.  

A variety of issues, some technical, have to be addressed in order to define 
eq.~\ref{d3first}\ precisely: (i) The integration measure for 
integrating over metrics is notoriously nonlinear; (ii) Solving the 
constraint $W=0,$ and then extracting the volume of the group of 
diffeomorphisms will leave a Jacobian; (iii) 
What is an appropriate choice for ${\cal L}$, given the  constraint $W=0?$ 

Before tackling the  technical issues,   consider the logic of what
we wish to do. Metrics with 
vanishing Weyl curvature are of the form 
\begin{equation}
h_{\alpha\beta}(x) = {{\partial f^{\delta}}\over {\partial x^{\alpha}}}
{{\partial f^{\epsilon}}\over {\partial x^{\beta}}}\ 
\ee{2\phi(f(x))}\ \delta_{\delta\epsilon} \equiv 
J^{\delta}_{\alpha}J^{\eps}_{\be}\ee{2\phi(f(x))}\ \delta_{\delta\epsilon}.
\end{equation}
For such metrics, $\sqrt{h}  = \det(J) \ee{d\phi}.$  Now, following 
our steps in Sect.~2, we want to extremize (ignoring technicalities) 
\begin{equation}
  \hat S \equiv \int d^{d}x \det(J) \ee{d\phi}
{\cal L}[\phi(f(x))] +i\lambda\left(\int d^{{d}}x \det(J) \ee{d\phi} - 1\right)
- N \int d^{{d}}x \hat P(x)\left[\ln\det J 
+\phi\right]\ ,
\label{d3action}
\end{equation}
where $N\hat P(x)\equiv\sum\delta(x-x_{i}).$  
Now, $\hat P(x)$ is a density, so $ d^{{d}}x \hat P(x)= d^{{d}}f \hat 
P(f),$ with $\hat P(f) \det(J) = \hat P(x).$  We can rewrite 
eq.~\ref{d3action}\ as
\begin{equation}
\hat S = \int d^{d}f  \ee{d\phi}
{\cal L}{[\phi(f)]} +i\lambda\left(\int d^{{d}}f   \ee{d\phi} - 1\right)
- N \int d^{{d}}x \hat P(x) \ln\det J 
 - N \int d^{{d}}f \hat P(f)\phi\ ,
\label{d3actionf}
\end{equation}
It follows that there are two
independent  functional variations, one with respect to $\phi(f),$ 
and the other with respect to $f(x),$ the latter giving
\begin{equation}
{\part\over{\part x^{\al}}}\left[ \hat P(x){{\partial x^{\alpha}}
\over {\partial f^{\delta}}}(x)\right] = 0.
\label{bu}
\end{equation}
Eq.~\ref{bu} is the analogue of eq.~\ref{acc2}, and
 determines the co\"ordinates $f^{\alpha}$
in terms of the co\"ordinates $x^{\alpha}$ in which the data is
presented, and is reparametrization-invariant because $P(x)$ is 
a density, not a function.  
Consider the special case $x=f:$   eq.~\ref{bu} is 
the requirement that the density $P$ in the co\"ordinates $f$ is the 
{\it constant} density, exactly analogous to our result in one 
dimension, with $\phi$ a constant.  

It is amusing to explicitly 
construct   $f^{\alpha}$ co\"ordinates which satisfy 
\[\det {{\partial f^{\alpha}}\over{\partial x^{\beta}}} = \hat P(x).
\]
We construct these co\"ordinates essentially iteratively as a 
one-dimensional distributions with parameters.  If we couldfind $f:$ 
\[f^{1}(x^{\alpha}) = f^{1}(x^{1}),\quad  f^{2}(x^{\alpha}) = f^{2}(x^{1}, 
x^{{2}}), \ldots \]
then we would get 
\[\det {{\partial f^{\alpha}}\over{\partial x^{\beta}}} =  
\prod_{\alpha} {{\partial f^{\alpha}}\over{\partial x^{\alpha}}},
\]
so the problem is one of factoring $\hat P$ appropriately.  Define
\begin{eqnarray}
&&P^{1}(x^{1}) \equiv  \int \hat P(x^{1},x^{2},x^{3},\ldots) 
 \prod_{\alpha=2}^{d} 
dx^{\alpha} ,\nonumber\\
&&P^{2}(x^{1},x^{2})P^{1}(x^{1}) \equiv  \int  
\hat P(x^{1},x^{2},x^{3},\ldots) \prod_{\alpha=3}^{d} 
dx^{\alpha}  , \nonumber\\
&&\vdots \nonumber\\
&& P^{d}(x^{\alpha}) \prod_{\beta=1}^{d-1} P^{\beta} \equiv 
\hat P(x^{\alpha}),
\end{eqnarray}
then 
\begin{eqnarray}
&&f^{1}(x^{1 })\equiv  \int^{x^{1}} P^{1}(z) dz ,\nonumber\\
&&f^{2}(x^{1 },x^{2})\equiv  \int^{x^{2}} P^{2}(x^{1},z) dz ,\nonumber\\ 
&&\vdots
\end{eqnarray}
is an explicit set of coordinates in which the given density $\hat P$
is constant.  Notice that each $f^{\alpha}$ varies from 0 to 1, as 
befits integrals of one-dimensional normalized probability 
distributions,  $P^{\alpha}.$

As explained in 
Ref.'s\cite{bialek,me}, we are interested in extracting a
continuum distribution from some set of binned  data,
 as independent as possible of 
the binning.  In renormalization group terminology, we are interested 
in the `infrared' properties of the data, independent of the 
`cutoff'.  The terms most relevant in the infrared are precisely the 
terms with the fewest derivatives. 
Thus, since  
Einstein-Hilbert action is a total derivative for conformally Euclidean 
metrics, we  
consider ${\cal L} \equiv - {1\over 2} R{\Delta}^{{-1}}R,$
where $\Delta$ is the Laplacian, and $R$ is the Ricci scalar 
constructed from $h_{\alpha\beta}.$  This Lagrange density is non-local 
for general metrics in dimensions larger than two, 
but is {\it local} for metrics with $W=0,$
\begin{equation}
\int d^{d}x\ \sqrt{h}\ {\cal L} = {(d-1)^{2}\over 2}\int d^{d}f\  
\ee{(d-2)\phi(f)} 
\delta^{\alpha\delta}
{{\partial \phi}\over {\partial f^{\alpha}}}
{{\partial \phi}\over {\partial f^{\delta}}}.
\end{equation}
The steps carried out in the previous section for determining the
finite $N$ corrections can be carried out in an exactly analogous 
fashion for this action for the leading correction, using the 
Laplacian in $d$ dimensions in the $f$ co\"ordinates explicitly
constructed above.  At higher orders in $N,$ the nonlinearity of 
$\cal L$ leads to a distinctly different computation.  I hope to 
address these issues, along with the technical issues mentioned above,
elsewhere. 

In conclusion, I have computed the finite sample size corrections to 
the geometric solution of the statistical inference problem I gave in
earlier work.  The uniform convergence of these corrections should be 
important for  applications so it is of interest to test this 
theoretical
construction in practice.

I am grateful to V. Balasubramanian and C. Callan for conversations. 
This work was supported in part by NSF grant PHY96-00258.

\end{document}